\documentclass[preprint,12pt]{aastex}
\usepackage{epsfig}
\begin{document}
\title{
Short-duration lensing events: I.~wide-orbit~planets?~free-floating 
low-mass objects? or high-velocity stars? 
}
\author{Rosanne Di\thinspace~Stefano}
\affil{Harvard-Smithsonian Center for Astrophysics, 60
Garden Street, Cambridge, MA 02138}


\begin{abstract}
Short duration lensing events tend to be generated by low-mass lenses
or by lenses with high transverse velocities.  
Furthermore, for any given lens mass and speed, events of
short duration are preferentially caused by nearby lenses (mesolenses) 
that can
be studied in detail, or else by lenses so close to the source star
that finite-source-size effects may be detected, yielding information about both
the Einstein ring radius and the surface of the lensed star.
    
Planets causing short-duration events may be in orbits 
with any orientation, and may have 
semimajor
axes smaller than an AU, or they may reach the outer limits 
of their planetary systems, in the region corresponding to the Solar System's
Oort Cloud. 
They can have masses larger than Jupiter's or smaller than Pluto's.
Lensing therefore has a unique potential to expand our understanding 
  of planetary systems. 
A particular advantage of lensing is that it
can provide precision measurements of system parameters, including
the masses of and projected separation between star and planet. 
We demonstrate how the parameters can be extracted and show that 
a great deal can be learned. For example, it is remarkable 
that the gravitational mass of nearby 
free-floating planet-mass lenses can be measured
by complementing observations of a photometric event
with deep images that detect the planet itself.

A fraction of short
 events may be caused by high-velocity stars
located within a kpc.
Many high-velocity lenses are likely to be neutron 
stars that received large natal kicks. Other high-speed
stars may be members of the halo population.  
Still others may be hypervelocity stars that have been 
ejected from the Galactic Center, or runaway stars escaped
from close 
binaries, possibly including the progenitor binaries of
 Type Ia supernovae. 
\end{abstract} 

\section{Introduction} 
\def\ev{event}

Typical lensing events are highly
degenerate: a given event may correspond to lensing
by any of a wide range of masses. (See Dominik 2009 and references therein.)
We demonstrate that, when the
Einstein-diameter crossing time, $\tau_E,$ of an event
is short enough, only a narrow range of physical
models are possible. 
For values of $\tau_E$ on the order of days
the lens can be a planet, a brown dwarf,
or a high-velocity star. 

The lens location is also constrained by the event duration.
If we consider a given type of lens, characterized by its
mass and a range of spatial velocities, the shortest events
are associated the lenses that lie very close to the observer, or else
very close to the lensed source. 
Nearby lenses are called mesolenses because the astrometric and photometric
effects of lensing both may be detectable, 
and because the possibility of observing the lens system directly
opens new avenues of study (Di\thinspace Stefano 2008a, 2008b).
This is particularly intriguing for planetary systems, because the
full range of available 
techniques (lensing, radial velocity studies, transits,
and direct detection of planets) may be applied 
to individual systems, making  them 
among the best-studied planetary systems. 
Similarly, lensing by
nearby brown dwarfs
can produce high-precision mass measurements.
In fact, two short events
caused by nearby brown-dwarf lenses have already been observed. (See
Fukui et al.\, 2007; 
Gaudi et al.\, 2008;
Gould et al.\, 2009).

Through the study of short events we can also learn more about 
high-velocity lenses, especially those that are nearby. 
Because the rate of events is proportional to
the angular speed, high-speed lenses should produce more events
per unit mass than slower-moving lenses. 
High velocities are expected for some classes of objects,
such as halo stars. 
Neutron stars are also promising because they can receive natal
kicks. We know that, within a kpc of Earth, 
there is a large population ($10^6-10^7$) of 
non-pulsing neutron stars, but only a relative handful have
been discovered (see, e.g., Haberl 2005). A significant fraction 
(a few percent; Di\thinspace Stefano 2008b) of detected 
lensing events are likely caused by nearby neutron stars, making
lensing events a potentially very productive way of discovering 
neutron stars and conducting mass measurements. Among short events, 
a significant fraction not generated by planets 
are likely to be generated by high-speed neutron stars 
(Di~Stefano 2009).

Hypervelocity stars, apparently ejected from the Galactic center, 
 are an exciting and relatively recent discovery  
(see Brown et al.\, 2009 and references therein).
While many, including the prototype, SDSS J090745.0+024507 (Brown
et al.\, 2005) are unbound, there is also a bound component
(Brown
et al.\, 2007). While the hypervelocity stars appear to have been
ejected from the Galaxy center (see, e.g., Ginsburg \& Loeb \ 2007; 
cf. Abadi et al. 2009), there is also
a population of high velocity ``runaway'' stars that appear to have
originated in the Galactic disk (Blaauw 1954, 1993; Bromley et al.\, 2009).
Lensing events are most likely to be
detected when the lens is dim; hypervelocity and runaway stars
are therefore most likely to produce detectable events after
becoming stellar remnants.
For example, some high-speed white dwarfs
may have started their lives in binaries
that later produced Type Ia supernovae. 
Unlike core collapse supernovae, Type Ia supernovae do not leave
a long-lasting remnant: the matter from a white dwarf that achieves
the Chandrasekhar mass undergoes explosive nuclear burning and is
dispersed. In a very promising class of models, however, the white dwarf
gains mass from a non-degenerate companion. The companion is not
destroyed by the explosion, and is free after it occurs. If the
binary was a close one, the speed of the companion star may be 
$\sim 200$~km~s$^{-1}$. Should close binaries constitute the 
major class of Type Ia supernova progenitors, then the Galaxy could 
house
$\sim 10^7$ high-speed stars, most of them old white dwarfs descended from 
companions to accreting white dwarfs. Although these are only a small fraction
of all white dwarfs, the rate at which they cause lensing
events can be significant because they have higher angular
speeds than other white dwarfs. 
In addition, the initial masses of the companion stars in 
Type Ia supernova progenitors may  
be large enough to produce white dwarfs with masses $> 0.6\, M_\odot,$
thereby also ensuring that these products of binary progenitors
produce lensing events at rates higher than other white dwarfs. 
The failure to find fast moving white dwarfs that are part of
a disk population could 
place constraints on Type Ia progenitor models.

Whatever the nature of the lens, if it
is not nearby, but is instead close to
the lensed source star, finite-source-size effects are more
likely to be 
detectable. Finite-source-size can be used to measure the  
Einstein angle,
$\theta_E$, of the lens, thus providing a relationship between
the lens mass and its distance, $D_L$
 from us. Together with the value of $\tau_E$
derived from a fit to the light curve, $\theta_E$ can be used to compute
 the value of the
angular speed $\omega=v/D_L.$ We can then test models in which
both $v$ and $D_L$ are relatively small, or relatively large. 
Finite-source-size effects also allow us to 
explore
features of the source star's surface.

In \S~2 we establish that an event with 
short Einstein-diameter crossing time can be caused only
by a limited set of lenses.
Section~3 is an overview of the types of measurements that can be
used to identify the 
correct physical model for the lens producing a short-duration
 event. We focus on effects other than caustic crossings,  
which were 
studied  for wide-orbit planets in Han 2005. 
We show that in many cases the lens mass and its distance from us
can be determined through a combination of (a)~studying the light curve,
(b)~determining the source size, (c)~measuring the astrometric effects
of lensing, and (d)~detecting the lens or placing limits
on the flux we receive from it. 
These same procedures can determine if
a planet orbits a star and, if it does, can measure the mass ratio
and projected orbital separation. 
Such a rich set of tests is available that independent
measurements of key quantities, such as the lens mass,
can be made in some cases. 
In \S~4, we explicitly consider the case in which
the lens is a free-floating planet and demonstrate that, 
especially for nearby planets, the model can be well tested and that mass
measurements may be possible in some cases.
\S 5 sketches the advantages of focusing attention on
events of short duration.

\section{Short-duration Events: What Can the Lenses Be?}

\subsection{Relevant Equations} 
 
If the mass of the lens is $M$ and its distance from us is $D_L,$
then the Einstein angle is: 
\begin{eqnarray} 
\theta_E &=& 9.025\,  {\rm milliarcsec}\, 
       \Big({{100\, {\rm pc}}\over{D_L}}\Big)^{{1}\over{2}}  
                    \Bigg[
                    \Big({{M}\over{M_\odot}}\Big)\, 
                    \Big({{1-x}}\Big)\Bigg]^{{1}\over{2}}\nonumber \\ 
 &=& 0.2788\,  {\rm milliarcsec}\, 
       \Big({{100\, {\rm pc}}\over{D_L}}\Big)^{{1}\over{2}}  
                    \Bigg[
                    \Big({{M}\over{M_{Jupiter}}}\Big)\, 
                    \Big({{1-x}}\Big)\Bigg]^{{1}\over{2}}
\end{eqnarray}
Here, $x=D_L/D_S,$ and $D_S$ is the distance to the lensed source.   
When the angular separation between the source and lens is
$\theta_E$ ($2\, \theta_E$, $3.5\, \theta_E$), the magnification
is $34\%$ ($6\%$, $1\%$).  

The relative angular speed between source and lens is $\omega.$
For nearby lenses, 
we have 
\begin{equation}
 \omega =
\Big({{0.0527''}\over{{\rm yr}}}\Big)\,
\Big({{100 \, {\rm pc}}\over{D_L}}\Big)\,
\Big({{v}\over{25\, {\rm km/s}}}\Big).  
\end{equation}
For lenses closer to the lensed source, the relative motion may not be
dominated by the motion of the lens. The expression for $\omega$ 
is more properly  represented as a sum of terms, but we can use Equation 2
to derive an approximate value by setting $v$ equal to the relative transverse speed.

Let $\tau_{E}$ represent the Einstein diameter crossing time.
\begin{eqnarray}
\tau_E &=& {{2\, \theta_E}\over{\omega}}\nonumber\\
 &=& 3.86\, {\rm days}\ \Bigg({{25\, {\rm km/s}}\over{v}}\Bigg)\, 
\Bigg[\Bigg({{D_L}\over{100\, {\rm pc}}}\Bigg)\, 
\Bigg({{M}\over{M_{Jupiter}}}\Bigg)\,
\Bigg(1-x\Bigg)\Bigg]^{{1}\over{2}}     
\end{eqnarray}
The actual duration of the deviation detected in any realistic
case depends on the distance of closest approach between the source and lens and
also on the photometric sensitivity. OGLE and MOA are both sensitive
to deviations on the order of $1\, \% $, so that the duration of the
detectable portion of events is generally longer than $\tau_{E}.$ 

Events 
in which light from the lensed source is
heavily blended with light from other stars 
can mimic short-duration events, even when the genuine  
Einstein diameter crossing times 
are weeks or months (Di~Stefano \& Esin 1995). 
The hypothesis that each short-duration event
is simply a highly blended event with a larger value of $\tau_E$
 must therefore be considered (see Di\thinspace Stefano \& Esin 1995;
Dominik 2009),
By tracking the event in several wavebands  
we can measure the fraction of the
baseline light contributed by the lensed source
in each, determining its 
color and spectral type. 
 In addition to quantifying the amount of blending, this
facilitates the identification of the lensed star in 
high-resolution images and allows us to estimate its radius.  

The mass of the lens can be expressed as follows. 

\begin{eqnarray} 
{M} &=& 
{{21.3\, M_\earth}\over{(1-x)}}\, 
\Big({{D_L}\over{100\, {\rm pc}}}\Big)\,  
\Big[\Big({{\tau_E}\over{1\, {\rm d}}}\Big)\, 
     \Big({{\omega}\over{0.0527^{\prime\prime}/{\rm yr}}}\Big)\Big]^2 \nonumber\\
&=& 
{{21.3\, M_\earth}\over{(1-x)}}\, 
\Big({{D_L}\over{100\, {\rm pc}}}\Big)\,  
\Big[{{\theta_E}\over{0.072 \, {\rm mas}}}\Big]^2  
\end{eqnarray} 

\noindent 
The value of $\tau_E$ can be determined from fitting the light curve. 
In many cases, the values of $D_L$ and $\omega$ (or $D_L$ and $\theta_E$)  
can be measured,
producing a high-precision measurement of the lens mass. When 
$D_L$ and $\omega$ cannot be measured, models of the Galaxy can be used
to derive probability distributions for them which can, in turn, be
used to construct a probability distribution for the lens mass.  

\subsection{Lenses Producing Short-Duration Events}

\subsubsection{Low-Mass Lenses}

Here we will make an empirical distinction between brown dwarfs and planets. 
Let $M_J$ represent the mass of Jupiter. 
Then, objects with $10\, M_J < M < 0.08\, M_\odot$ will be referred to
as brown dwarfs and objects of lower mass will be referred to as planets.
Those planet-mass objects that are not bound to stars will be referred
to as free-floating planets. 

Consider lenses that are part of the Milky Way's disk or halo
population. Values of the transverse
speed may range from roughly $10$~km~s$^{-1}$ to $200$~km~s$^{-1}.$
Figure 1 illustrates that an event with small $\tau_E$ can only be generated
by a low-mass lens, either a planet or a brown dwarf.
The left panel 
was generated for a transverse speed of $75$~km~s$^{-1}.$
Consider a short event 
with a light-curve fit yielding a value of $\tau_E$.
The swath corresponding to this value of $\tau_E$ shows the range of
possible lens masses, M, and distances from us, $D_L$.
For example, given a transverse speed of $75$~km~s$^{-1},$
events with $\tau_E<1$~day can only be
generated by planet-mass lenses. 
Even events with durations near $4$~days
can be produced by objects as massive as 
brown dwarfs only if the lens lies
within a roughly a hundred pc
of us, or a similar distance from the Galactic center.
Because relatively few brown dwarfs lie within these 
small volumes, the rate of very short brown-dwarf-lens events
is limited. Those very short events that are generated by
brown dwarfs are very interesting, however, since the lens
will be detectable if it is nearby, and will produce detectable
finite-source-size effects if it is in the source system. 

Many nearby lenses have smaller transverse
speeds (see, e.g. L\'epine 2005), while halo stars would likely have
larger transverse
speeds (see, e.g., Dominik 2006).
The left panel of Figure~1 can be easily adapted for any value of $v$:
when the transverse
speed is $75$~km~s$^{-1}/N,$ then the values of $\tau_E$ for each swath
should be multiplied by $N$.
For the slower (faster)
speeds, any given time duration corresponds to a lens of lower
(higher) mass.

\subsubsection{High-velocity Lenses} 

A combination of small (or large) $D_L$ and large $v$ 
can conspire to produce short lensing events.
(See the right panel of Figure 1.)
Stars moving with velocities  above several hundred
km~s$^{-1}$ are rare,
but they do exist (\S 1). The two most well-known examples are neutron stars
which receive strong natal kicks (see, e.g., Chatterjee et al.\, 2009), 
and hypervelocity stars (see Brown et al.\, 2009 and references therein).
In addition, ``runaway stars'' which may have been ejected from clusters,
supernova-progenitor binaries or other close binaries, can have
velocities in excess of $200$~km~s$^{-1}$. (See, e.g., 
Hoogerwerf, de Bruijne, \& de Zeeuw\ 2001, and
references therein.)  

The lenses used to generate the points in the right panel of 
Figure~1 have masses of $1.4\, M_\odot,$
comparable to typical neutron star masses. They differ from each other in
their distances $D_L$ from us and in their transverse velocities, $v.$  
Each swath plotted in Figure~1 corresponds to
a specific value of $\tau_{E}.$ 
The right panel of Figure~1 shows that with a lens mass of $1.4\, M_\odot,$
$v$ must be greater than roughly $1000$~km~s$^{-1}$ 
($300$~km~s$^{-1}$) in order for the
Einstein-diameter crossing time to be $1$~day ($2$~days) or shorter. 
Even $4$-day events generated by high-velocity stellar-mass lenses
are likely to be rare, since for $\sim v=300$~km~s$^{-1}$,
the lens  must be within $\sim 20$~pc of Earth, or comparably close
to the lensed source.  

The right panel of Figure~1 can be easily adapted for any value of $M$:
when the lens mass 
 is $N\,\times  1.4\, M_\odot$, then the values of $\tau_E$ for each swath
should be multiplied by $\sqrt{N}$.

\subsubsection{Summary}  

In summary, Figure 1 illustrates that selecting for small values of
$\tau_E$ is guaranteed to select events with interesting lenses.
The shorter the event, the lower the mass of the lens and/or the faster
its transverse speed. 

Assuming that the local population of lenses and
the population in the source field is similar, roughly half of the 
short events are likely to be generated by lenses close enough to us
to allow detailed model tests. A significant portion of the 
more distant lenses may produce finite-source-size effects,
providing constraints on the lens and/or the lensed source.   
On the other hand, the nearby lenses can be studied by a variety
of complementary observations that can combine with measurements of
the photometric event to constrain the lens model and in some cases to
obtain a measurement of the gravitational mass of the lens and/or of
any stellar companion it may orbit.

\section{Overview of Model Tests: Breaking the Degeneracy} 

For each short-duration event, we would like
to answer several key questions. 
What is the nature of the lens? 
Is it a planet? If so, is
it free-floating or bound to a star? Is the lens a brown dwarf?
Is the lens a high-velocity object? If it is a high-velocity
object, what is it and from where
did it come? 
These questions can be answered by constraining the lens mass,
testing for evidence that the lens has a companion, and in some cases,
through the direct detection of the lens system. 
Observing teams are already sampling some short-duration events
with the photometric sensitivity and cadence needed to allow meaningful
model fits. 
Especially if alerts are called to encourage almost continuous
monitoring, the light curves will quantify the effects of blending,
and identify those events with genuinely short Einstein diameter
crossing times. Fits of the light curve will  
provide values of $\tau_E$ and $b,$ the angle of closest approach,
expressed in units of $\theta_E.$  They will also determine if there
are signatures of finite-source-size effects or other deviations
from the standard point-source/point-lens form.

\subsection{Determining $\theta_E$} 

When the angle of closest approach between the source and lens, 
$b\, \theta_E$, is comparable to the angular dimensions of the 
star, the light curve shape is altered (Witt \& Mao 1994; Lee et al.\, 2009 and references therein). 
Thus, when the Einstein angle is small, as is the case for low-mass
lenses at moderate to large distances,  
finite-source-size effects are most likely to be detectable through
their alteration of the light curve shape.
To set the scale, note that 
 $\theta_E$ for an Earth-mass
planet at $100$~pc is $1.58 \times 10^{-5}$~arcsec,
only $\sim 2.7$ times larger than the angular size of
a $10\, R_\sun$ star at $8$~kpc.  

When, on the other hand, the Einstein angle is
large it may be possible to measure  
the astrometric effects of lensing.  
The Einstein angle of a neutron star, for example, is 
$0.01^{\prime\prime}$
at $D_L = 100$~pc.\footnote{Neutron 
stars at roughly this distance are known (Haberl 2005),
but, with $\sim 7$ out of roughly $10^7$ nearby isolated non-pulsing 
neutron stars discovered so far,  
our census of nearby neutron stars is woefully incomplete.}   
Under ideal circumstances, this is measurable, because
the maximum shift in the source's centroid
of light is nearly equal to the size of the Einstein ring.
The maximum shift is attained when the lens-source separation is
approximately equal to $\theta_E.$ The magnitude of the shift
falls off as the inverse of the separation. 
(See Miyamoto \& Yoshii 1995; 
Hog et al.\, 1995; 
Miralda-Escude 1996; Dominik \& Sahu 2000;
 An \& Han 2002; Han 2002; Asada 2003; Han 2006.)   
To detect the shift, we need
good angular
resolution, a bright source, and a lens that does not dominate
in the wavelength range in which the source is most luminous.
Among short-duration events, 
those caused by high-speed stellar remnants provide 
the greatest opportunity to measure centroid shifts. 
Not only are their Einstein rings typically bigger, but in addition, 
isolated compact objects can be dim
(see, e.g., Haberl 2005), 
making it easier to measure the position of the
source star's centroid of light.  

An interesting case to consider is one in which a short-duration
event is caused by a planet lens that happens to be in orbit 
around a star.  
At the time of the short-duration photometric event caused by  
  the planet, the magnification of the lensed source is dominated by
the gravitational influence of the planet; the central star
has only a small influence on the magnification if the angular
separation, $\phi,$ between the lensed source and the central star 
of the planetary system is larger than
$\sim 1.5\, \theta_{E,\ast}.$ The Einstein ring of the planet is small
enough that present-day observations would not be 
able to detect the astrometric shift in the source position
 produced by the planet.  
It is possible, however, that at the time of the short event,
 the {\it central star} can produce a
detectable 
astrometric shift in the lensed source's light centroid. 
Depending on the size of its Einstein ring and
its luminosity relative to that of the lensed star, its astrometric
influence may be detected for $a$ as large as $\sim 10\, \theta_{E,\ast}$
using HST, and much farther using astrometry possible in future missions
(Han 2006). The detection of the astrometric effects associated
with lensing by the central star
can measure the value of its Einstein angle, $\theta_{E,\ast}.$ 

\subsection{Determining $\omega$ and $D_L$}

As a larger portion of the sky is monitored
with good photometric sensitivity, and as monitoring programs establish
longer baselines, the proper motions and parallaxes
of a large fraction of nearby stars and brown dwarfs 
will be measured and published. 
It will increasingly become the case that no new
observations are needed to establish that the lens is likely to be a nearby
low-mass object, and to measure its proper motion, and 
and its photometric and geometric parallax. In such cases, the mass of the lens can be determined
simply by using the light curve to measure $\tau_E$.\footnote{ There is a complication that must be addressed
when the lens is a planet orbiting a star, because the catalogued proper motion is  that of the star.
In such cases, however, it may be possible to determine or constrain the planet's orbit (\S 3.3, \S 5.3),
allowing us to determine the angular speed 
of the planet at the time of the event.}

Below we concentrate on  determining the values of $D_L$ and $\omega$
in the more difficult cases in which 
 little or nothing is known about the
lens at the time of the event.  
An example is a case in which the lens system is extremely
dim. Lenses that may not be detected even with
deep follow-up observations
include low-mass free-floating planets and 
high-velocity isolated black holes.
For other dim lens systems, there may be enough flux for detection, but
high-precision astrometry to measure proper motion and parallax may
be difficult. Fortunately, even when the lens is not detectable,
there are situations in which its parallax and/or proper motion can be 
measured. 

First, parallax effects can influence the shape of the light curve, even for short events. 
This was illustrated by a recent short-duration event 
in which the lens was a brown dwarf located $525\pm 40$~pc away 
(Gould et al. 2009).
The lens mass is $M=0.056\pm 0.004\, M_\odot,$
and its transverse velocity is $113\pm 21$~km~s$^{-1}$. The degeneracy
was broken through a combination of finite-source-size and 
parallax effects.
The reason parallax effects were detectable in this short event was that
the magnification was extremely high; in fact it was the high magnification
that caused an alert to be called (Griest \& Safizadeh 1998). 
Even with a less extreme 
magnification, however, a combination of 
other effects can sometimes accomplish the same goal.

In some cases
the proper motion can be measured
even without direct detection of the lens.
When, for example,  
the lens is a high-proper-motion object moving across a dense stellar field,
it may generate a sequence of independent events. In each, a different
source star is lensed. The angular distance between sources, combined
with the time between events, then measures the angular speed of the lens.  
This method relies only on measuring the
angular separation between source stars that were lensed at known
times. 
For some  nearby lenses, i.e., mesolenses,
 the angular speed can
be high enough, and the Einstein rings large enough that sequential
events are expected. A free-floating planet with $D_L=10$~pc
 with a transverse
speed of $50$~km~s$^{-1}$ covers an arcsecond per year;
a neutron star at $100$~pc with a speed of $200$~km~s$^{-1}$ 
covers $0.4$ arcsecond per year.\footnote{The 
rates generally depend on the background density of sources 
and 
the value of $\theta_E.$ For some nearby lenses, sequences
 of low-magnification
events are expected to occur over an interval of years to decades
(Di~Stefano 2008a.)}   

When the lens itself can be detected, as may be
the case for nearby brown dwarfs,
then   
we can measure $\omega$ directly. Because the lens and source are
coincident at the time of the event, a single image taken  
after they can be resolved will determine the 
value of $\omega$ and also
the photometric parallax of the lens, yielding the value of $D_L.$
A single WFPC2 
image taken $\sim 6$ years after the MACHO LMC-5 event
obtained a very good 
measurement of both $\omega$ and $D_L$ (Alcock et al.\, 2001)
\footnote{The lens in this case was an M~dwarf, not a brown dwarf.}.
Additional measurements and analyses confirmed this, and provided 
a high-precision mass measurement of an isolated M~dwarf 
(Gould 2004;
Gould, Bennett,
\& Alves 2004; Drake, Cook, \& Keller 2004; Nguyen et al.\, 2004). 

Consider the case in which the lens is a planet
orbiting a star that can be detected.
In this case, it may be possible to measure the distance to the star
and its angular speed.  
The distance $D_L$ 
to the planet is approximately the same as the distance to the
star. The angular speed of the planet is likely to be roughly equal to the
angular speed of the star for wide orbits. To refine
the measurement of the planet's proper motion at the 
time of the event, we must estimate the 
amount by which the two angular speeds differ. 
One way to estimate the difference is to measure the projected
separation, $a,$ between the planet and star (see \S 3.3 and \S 5.1). 

Finally, even if the lens system is not detected and even if it
generates just one event, 
we can
determine the value of $\omega$
if both $\tau_E$ and $\theta_E$ have been measured: 
\begin{equation} 
\omega={{2\, \theta_E}\over{\tau_E}}=
{{0.0114^{\prime\prime}}\over{{\rm year}}}\, 
\Bigg({{\theta_E}\over{{1.56 \times 10^{-5}}^{\prime\prime}}}\Bigg)
\Bigg({{1\, {\rm day}}\over{\tau_E}}\Bigg).
\end{equation} 
 
\subsection{Determining System Parameters}

We would like to determine the intrinsic properties of the lens, including
 its mass,
and 
whether or not it has companions, 
As the example of MACHO-LMC-5 illustrates, the mass of a lens can be
measured if $\tau_E, D_L,$ and $\omega$ are measured.
The mass of the lens can be the most significant clue to its nature.
If, for example, we find that the mass of a nearby lens
 is $0.001\, M_\odot, 1.4\, M_\odot,
7.0\, M_\odot$, and there is no obvious optical counterpart, then the    
lens is likely to be, respectively, a free-floating 
planet, neutron star, black hole.

Determining whether or not the lens generating a short-duration event
 has a companion 
can be simple or challenging, depending on the situation. 
Here we will consider the case in which the lens is a planet 
which orbits a star. 
The Einstein angle of the star can be written as follows.  
\begin{equation}  
 R_{E,\ast} =   0.4513\, {\rm AU} \times \Bigg[
    \Big({{M}\over{0.25\, M_\odot}}\Big)\, 
    \Big({{D_L}\over{100\, {\rm pc}}}\Big)\, 
    \Big(1-x \Big)\Bigg]  
\end{equation}

In the ``classical'' events which have been used to discover planets with lensing,
the projected orbital separation, $a$, lies between $0.5\,  R_{E,\ast}$
and $1.5\,  R_{E,\ast}.$  This is 
the zone for resonant  lensing, so named  because the 
planet-induced deviations are often associated with caustic crossings
and can be dramatic. 
For planets in the zone for resonant lensing
the 
 event is not short. Instead, the time duration of the event is
determined by the Einstein-diameter crossing time of the central star,
while evidence of the planet is provided by a short-lived deviation 
from the stellar-lens event.
For both smaller (Di~Stefano \& Night 2008)  and larger separations (Di~Stefano \& Scalzo 1999a), 
isolated events of short-duration are
expected. 
When $a < 0.5\,  R_{E,\ast}$, the short-duration events should  
all 
bear the signature of the star's influence. 
This range of separations is
interesting, corresponding to the habitable zones of many nearby 
stars (Di~Stefano \& Night 2008). 
Nevertheless, in the remainder of the paper we focus  
on planets in wide orbits ($a>1.5\,  R_{E,\ast}$).  
These  wide-orbit planets alone 
they are likely to produce more events than planets  
in the zone for resonant lensing (see also Di~Stefano \& Scalzo 1999a, 1999b).
We return to the short-duration events expected when the orbital separation
is small in the 
companion paper (Di~Stefano 2009).

For separations in the range between roughly $1.5\, \theta_{E,\ast}$~and 
$3.5\, \theta_{E,\ast},$ the gravitational influence of the 
central star is significant enough to
 perturb the short-duration planet-lens
light curve 
from the point-lens form that would have been created by a
free-floating planet. (See Figure 2.)  
The star-induced 
deviation 
is long-lasting, because its duration is comparable to the Einstein-diameter
crossing time of the star. It therefore frames the planet-lens light curve.
This star-induced deviation 
(a) can be used as a clue that the lens is a planetary system; when a
shorter higher-magnification event occurs on top of it, observers can
call alerts with a high degree of confidence that the lens is a planetary
system; (b)  
can be used to
obtain the possible values of the mass ratio ($q=m_{planet}/M_\ast$),
and also the orbital separation, $a,$ 
expressed in units 
of the Einstein radius of either
the star or planet.  
As mentioned in section~3.1, 
it  may be possible to measure astrometric shifts associated with 
lensing by the central star and to thereby measure the star's Einstein angle.
Generally such measurements would require HST. It is therefore fortunate
that, when the 
separation is in the range above, there is also a photometric signature
that provides clear evidence for the possible value of 
HST observations.

In addition,
for all values of $a$ large enough that the planet causes an independent event,
there are sets of source tracks that pass behind the lensing regions of both
the planet and the star it orbits. 
If the track of the source passes near both the planet and star,
two independent \ev s are generated, 
one of short duration and the other of longer
duration. For such repeating events, the values of $q$ and $a$ can be
derived (Di~Stefano \& Mao\ 1996;
Di~Stefano \& Scalzo 1999b).  

Note that none of the information mentioned in the paragraphs above relies on
detecting the central star. Therefore, even in the case in which the central 
star is a dark compact object, a neutron star or black hole, 
it may be possible to derive a great deal of
information about the planetary system.
In many cases, however, particularly if the system is nearby, the star can be
detected. This makes it possible to learn about the system
in a variety of other ways. For very wide orbits, the separation between the
central star and lensed star (hence the planet) may be measurable at the
time of the event. Whatever the orbital separation, detection
of the central star can provide an opportunity to directly measure 
$\omega$ and $D_L.$ 

The combination of information derived through lensing and 
direct detection of the  
central star can provide a detailed model of the planet-star
system. In addition, if the system is close enough, subsequent
studies can be carried out to search for radial velocity variations,
transits, and even to conduct imaging studies
(Marois et al.\, 2008, Kalas et al.\, 2008).  
 In some cases these could detect evidence of
the planet that produced the short-duration event. In others, 
additional planets can be discovered.    

\section{Planets: Bound or Free-Floating?} 

Our understanding of the formation of low-mass objects 
is still evolving.
In particular we don't know the mass distribution of
 low-mass
objects bound to stars, or the frequency with which 
such  objects are ejected. Nor do we know
the low-mass limit for objects that can form in isolation. 
 (See, e.g., Kroupa \& Bouvier 2003; Gahm \& Kristen 2007;
Hurley \& Shara 2002).
It is therefore difficult to predict
the numbers and characteristics of
free-floating planets.
Observations are difficult as well, because the planets would be
small, cool, and dim. 
As lenses, however, free-floating planets
produce effects that are the same as the effects produced by
planets orbiting stars in wide orbits. 

For planets orbiting stars, the detection of the star can provide important
information such as a value for $D_L$ and an estimate of the transverse
speed. For free-floating planets, no such source of additional information
is available. In fact, the absence of a stellar companion can be 
taken as indirect evidence that a lens producing a short-duration
event is free-floating, but only if we can establish that the lens is close
enough that any stellar companion would be detectable. 
Fortunately, the combination of short event duration and the ability
to detect or place limits on finite-source-size effects is
very powerful. 

Suppose that a combination of studying the light curve and identifying the 
lensed source has allowed us to either measure the Einstein angle
or to place a strong upper limit: $\theta_E\geq \theta_E^{lim}.$ 
Combining equations (2) and (5), we can express $D_L$ as follows.  
\begin{equation} 
D_L \leq 463\, {\rm pc}\, \Bigg({{v}\over{25\, {\rm km/s}}}\Bigg)\, 
                      \Bigg({{\tau_E}\over{1\, {\rm day}}}\Bigg)\, 
   \Bigg({{{1.56\times 10^{-5}}^{\prime\prime}}\over{\theta_E^{lim}}}\Bigg).   
\end{equation} 

The measurement of $\tau_E$ and $\theta_E^{lim}$, with values in the 
range considered above, accomplishes three things. 
First, as the relation above shows, these measurements can establish that the 
lens is nearby. The only caveat is that the transverse velocity should be
in the range of velocities typical of either disk stars or halo stars.  
Even for extreme stellar velocities, the derived limit can be
important. For example if the lens is 
a halo object with a transverse
velocity of $250$~km~s$^{-1}$, it must lie within $5$~kpc.
The smaller the value of $D_L,$ the more  likely it is that
we can detect a stellar companion, if one exists.  
  Second, 
measurement of $\tau_E$ and $\theta_E^{lim}$ 
establishes that the mass of the lens is likely to be
in the planet range. This can be seen by combining Equations (4) and (7).
Third, measurement of $\tau_E$ and $\theta_E^{lim}$ 
 places a lower limit 
on the value of $\omega$ [Equation (5)]:  
$\omega \ge  0.0114^{\prime\prime}/yr\, 
[\theta_E^{lim}/(1.56\times 10^{-5})^{\prime\prime}]\,
[(1 {\rm day})/\tau_E].$

\subsection{Eliminating Bound-Planet Models}

Suppose that the lens is a planet orbiting a star. Let the
angular separation between the planet lens and the star it orbits
be $\eta^{\prime\prime}.$ 
The projected separation between them is 
\begin{equation} 
a=100\, {\rm AU}\, \eta^{\prime\prime} {{D_L}\over{100~{\rm pc}}}  
\end{equation} 
For large separations, the angular speeds 
of the planet and its central star will
be almost identical, so that $\omega=(2\, \theta_E)/\tau_E$
should be a good approximation to
the value of the star's proper motion.  
By looking for all high-proper motion stars within any
specified angle of the
event location, we can 
check for the presence of 
central stars with angular separations within a corresponding
range of values of $a.$ If such a star is catalogued or
found through new observations, the probability
is large that it is associated with the planet lens. In this case, the
distance to the lens can be estimated by the photometric or
geometric parallax of the associated star, and the planet's
 mass can be estimated.

If no such star is discovered, then it is still possible that the
planet orbits a star that is close enough to it that it cannot be
resolved from the lensed source at the time of the event. 
In this case the light from the central star would be blended
with light from the lensed source. Testing for blending in different
wavebands during the event would then determine the amount of light that
could possibly emanate from any star associated with the lens.
A high-resolution image taken soon after the event would then
both identify the lensed source and 
any other stars in the field. Additional images would determine if any stars
in the region have the requisite proper motion. 
(In addition, as discussed in \S 3, high-resolution images may also
be able to measure astrometric shifts in the source position caused by
a central star.)  
Even if high-resolution
images 
cannot be taken, however, images at later times
can test for the 
presence of a central star as the lens system moves away from the
event position.

The bottom line is that it is possible to place quantifiable limits
on the presence of a central star. These limits can be expressed 
in terms of the   
mass of the star, the orbital separation of the planet 
from it, and the transverse velocity of the lens.

\subsection{Direct Verification}

At the time of a planet-lens event, the planet and lensed star are
nearly coincident. After a time that can be as short as several years,
the planet may have moved far enough from the event location, that,
it may be possible to detect it.
If the majority of free-floating planets have masses comparable to
the Solar System's terrestrial planets, it is unlikely that 
they will be detected in the foreseeable future.
If they are more massive, however, direct detection will be possible
for some of them.
The feasibility of direct detection has been established by recent detections 
of wide-orbit planets.   
For example, three planets with masses between $5$~and~$13$ times the mass of
Jupiter have been discovered orbiting the star HR~8799, which
 is 39.4~pc away (Marois et al.\, 2008). 
These planets have projected distances from the star ranging from
$24$ to $68$~AU. At these distances the 
radiation they receive from the star contributes little to their
luminosity, which is instead dominated by cooling. 
In addition, a planet with 
mass estimated
to be no greater than a few times that of Jupiter has been discovered in a
wide orbit (119 AU) around a star, Formalhaut, which is
 $7.7$~pc away (Kalas et al.\, 2008). 
Planets 
with radii comparable to or larger than that of Jupiter, and 
temperatures of $\sim 1000$~K, can be        
imaged if their distances from us are not too great.
Imaging planets, even in these wide orbits, requires deep observations
which can only be undertaken if there is a good chance of successfully
finding a planet or else placing meaningful limits. For planets
orbiting stars, such as those around HR~8799 or Formalhaut, the star
itself provides a well-defined region within
which to conduct a search. For free-floating planets,
the occurrence of a lensing event provides a location about which to
search. If $\tau_E$ and $\theta_E$ are both measured, then
$\omega$ is known, hence the angular separation of the lens from the 
lensed source can be computed for all later times. 
Observers can then plan observations to occur several years after the 
event, in order to image the lens. This procedure 
 will image a subset of nearby free-floating planets. 

If enough photons are collected to estimate the planet's temperature,
then the range of possible radii will provide a range of possible
luminosities, allowing the distance $D_L$
(and the uncertainty in its value) to be estimated. In addition, 
a sequence of images may be able to establish the geometric 
parallax. Armed with an estimate of $D_L,$ the lens mass
can be determined in those cases in which $\theta_E$ has been measured 
through finite-source-size effects.\footnote{ Note that the analogous
 procedure is
simpler for planet lenses which are in wide orbits around a star
that is close enough to be detected. In the latter case, the 
photometric and or geometric parallax of the star provides 
the distance to the
star, which is, to high precision, equal to $D_L$.}  

\subsection{Eliminating High-Velocity-Star Models}

To distinguish between models in which the lens is a 
planet bound to 
a star or else is a free-floating planet, 
we search for evidence of a star that could be associated with the 
lens.
Even if a star is not detected, however, it is possible that the event
was caused by a high-velocity stellar-mass object rather than by a
free-floating planet. 
Consider a short event (say $\tau_E=1$~day), and a measured small Einstein 
angle (say ${\theta_E=1.56\times 10^{-5}}^{\prime\prime}$).
If the lens is not a planet, how fast 
must the lens be moving and how far away
must it be?  For example, with $D_S=8$~kpc, a neutron star would have
to have $v=435$~km/s, and would have to be located within
roughly a pc   
of the source star.
This is so unlikely, compared to the probability
of a stellar-speed planet-mass object within a few hundred pc,
that it can be effectively ruled out. In fact, if free-floating
planets are common, we will have dozens to hundreds of similar cases, 
and the hypothesis that the majority of the events they generate were
actually generated by high-velocity stellar
remnants can be eliminated. 

Note that the arguments above apply to the case in which the
Einstein angle has been measured and is small. If the light curve shape 
provides only an upper limit on the value of $\theta_E,$ then it is still
possible that the lens was a high-velocity object. High-resolution
images that check for a shift in the source's light centroid
can measure $\theta_E,$  if the lens is indeed a nearby 
stellar remnant. Alternatively, they can place an upper limit
on the value of $\theta_E,$ thereby at least providing a constraint on
stellar-mass models for the lens.

\section{Focus on Short Events} 

When an event with a small measured value of $\tau_E$ occurs, we are
guaranteed that the lens is of special interest.  
It could be a low-mass object such as a brown dwarf,  a planet bound to
a star, or a  
free-floating planet. Or it could be a high-velocity object, either a dim 
member of the Galactic halo or else a high-velocity stellar 
remnant, a product of an earlier high-energy event.  

Because short events are being discovered by current monitoring teams 
(Udalski 2003; Bond et al.\, 2001), it is possible to
initiate programs designed to learn about both low-mass and high-speed
lenses.  The science returns will be
 high, because each  class of lenses generating short events 
is important, each for its own reasons. 

Furthermore, the lenses will be nearby (within about a kpc) for
a large fraction
of short events.
This means that they are  amenable to a variety of complementary studies.
Short events could, for example, provide a unique way to discover
local fast-moving neutron stars which would otherwise be known only as
dim, unexceptional sources. Short events will almost certainly identify 
planetary systems which can be studied in a variety of ways. Some of
these will become the ``gold standards'' of their class, because 
key quantities such as the gravitational mass of the star and planet 
can be measured, sometimes in complementary ways 
(Di\thinspace Stefano 2009). 

We have shown that even short events generated by free-floating planets can be
correctly identified, and planet properties can be measured.
It is significant that this can be accomplished without
sophisticated astrometry missions
(cf. Han 2006). Our result relies on careful study of short event
light curves, including multiwavelength monitoring to identify the
effects of blending and finite-source size. We also require 
comprehensive catalogs and/or additional high-quality observations
of the region in which the event occurred to search for counterparts,
sometimes during and sometimes years after the event.
Lensing will even provide 
gravitational mass measurements
for some free-floating planets.

In the companion paper we demonstrate that the expected rate of
short events is high enough that programs designed to study them
will be productive. We also suggest a set of procedures designed
to optimize their efficacy. 
These procedures will not only identify individual interesting lenses, 
but will also allow population properties to be measured. 
We will be able to, for example, estimate the frequency of free-floating
planets,
both in absolute terms and also relative to the frequency of
planets orbiting stars. 
We will quantify the size of the population local neutron stars 
and their velocity distribution. 
We will discover or place limits on the
companions to accreting white dwarfs, flung from the vicinity of
Type Ia explosions, and we will discover or place limits on the number
of remnants of hypervelocity and runaway stars still bound to
the Milky Way.

\bigskip
\bigskip
\noindent {\bf Acknowledgments:} This work benefited from discussions
with  
Charles Alcock, Erin Arai, Mary Davies, 
Nitya Kallivayalil, M.J. Lehner, Christopher Night, Brandon Patel, 
Frank Primini, Pavlos Protopapas, Rohini Shivamoggi,  
Kailash Sahu, and Takahiro Sumi.  
The research was conducted under the aegis of an NSF grant,
AST-0708924, and 
a grant from the Smithsonian endowment. I would like to thank the
Aspen Center for Physics for its hospitality during the early
phases of this work.  

\bigskip
 
\centerline{\sl References} 

\noindent Abadi, M.~G., Navarro, 
J.~F., \& Steinmetz, M.\ 2009, \apjl, 691, L63 

\noindent
Alcock, C., et al.\ 
2001, \nat, 414, 617 




\noindent 
An, J.~H., \& Han, C.\ 2002, \apj, 573, 351 


\noindent 
 Asada, H.\ 2003, The Proceedings 
of the IAU 8th Asian-Pacific Regional Meeting, Volume I, 289, 441 
A Perturbative Approach to Astrometric Microlensing due to an Extrasolar Planet

\noindent 

\noindent

\noindent
Blaauw, A.\ 1993, Massive 
Stars:  Their Lives in the Interstellar Medium, 35, 207 

\noindent
 Blaauw, A., \& Morgan, W.~W.\ 1954, \apj, 119, 625 

\noindent

\noindent

\noindent
Bond, I.~A., et al.\ 2001, MNRAS, 327, 868
https://it019909.massey.ac.nz/moa/

\noindent
Bromley, B.~C., Kenyon, 
S.~J., Brown, W.~R., \& Geller, M.~J.\ 2009, \apj, 706, 925 

\noindent
Brown, W.~R., Geller, 
M.~J., \& Kenyon, S.~J.\ 2009, \apj, 690, 1639 

\noindent
Brown, W.~R., Geller, 
M.~J., Kenyon, S.~J., 
Kurtz, M.~J., \& Bromley, B.~C.\ 2007, \apj, 660, 311 

\noindent
Brown, W.~R., Geller, 
M.~J., Kenyon, S.~J., \& Kurtz, M.~J.\ 2005, \apjl, 622, L33 

\noindent

\noindent

\noindent
Chatterjee, S.\, et al.\, 2009, arXiv:0901.1436
 
\noindent
Di Stefano, R.\, 2009, {\it
Short-duration lensing events: II. Expectations and Protocols}, in preparation.

\noindent
Di Stefano, R.\, 2008a, ApJ, 684, 46

\noindent
Di Stefano, R.\, 2008b, ApJ, 684, 59

\noindent
Di Stefano, R., \& Esin, A.~A.\ 1995, \apjl, 448, L1

\noindent
Di Stefano, R., \& Mao, S.  1996, ApJ, 457, 93

\noindent
Di Stefano, R., \& Scalzo, R.~A.\ 1999b, \apj, 512, 579 

\noindent
Di Stefano, R., \& Scalzo, R.~A.\ 1999a, \apj, 512, 564 

\noindent
Di Stefano, R. \& Night, C.\, 2008, ArXiv e-prints, 0801.1510

\noindent
Dominik, M.\ 2009, \mnras, 
393, 816

\noindent
Dominik, M., \& Sahu, 
K.~C.\ 2000, \apj, 534, 213 

\noindent

\noindent 
Drake, A.~J., Cook, 
K.~H., \& Keller, S.~C.\ 2004, ApJL, 607, L29 

\noindent
Fukui, A., et al.\ 2007, \apj, 670, 423

\noindent
Gaudi, B.~S., et al.\ 
2008, \apj, 677, 1268


\noindent
 Ginsburg, I., \& Loeb, A.\ 2007, \mnras, 376, 492

\noindent
Gould, A., et al.\ 2009, 
arXiv:0904.0249

\noindent
Gould, A., Bennett, 
D.~P., \& Alves, D.~R.\ 2004, ApJ, 614, 404 

\noindent Gould, A.\ 2004, ApJ, 606, 319 


\noindent 
Griest, K., \& Safizadeh, N.\ 1998, \apj, 500, 37

\noindent
Haberl, F.\ 2005, 5 years of 
Science with XMM-Newton, 39 




\noindent
Han, C.\ 2006, \apj, 644, 1232 

\noindent
Han, C.\ 2005, \apj, 629, 1102 

\noindent
 Han, C.\ 2002, \mnras, 335, 189 


\noindent
Hog, E., Novikov, I.~D., \& Polnarev, A.~G.\ 1995, \aap, 294, 287 

\noindent
Hoogerwerf, R., de Bruijne, J.~H.~J., \& de Zeeuw, P.~T.\ 2001, \aap, 365, 49 


\noindent
Kalas, P., et al.\ 2008, 
Science, 322, 1345 



\noindent
Lee, C.-H., Riffeser, A., 
Seitz, S., \& Bender, R.\ 2009, \apj, 695, 200 

\noindent
L{\'e}pine, S.\ 2005, \aj, 
130, 1680


\noindent
Marois, C., Macintosh, 
B., Barman, T., Zuckerman, B., Song, I., Patience, J., Lafreni{\`e}re, D., 
\& Doyon, R.\ 2008, Science, 322, 1348 

\noindent
Miralda-Escude, J.\ 
1996, \apjl, 470, L113 

\noindent
Miyamoto, M., \& Yoshii, Y.\ 1995, \aj, 110, 1427 

\noindent
Nguyen, H.~T.et al.\,
2004, ApJS, 154, 266





\noindent
Udalski, A. \, 2003, Acta Astron., 53, 291


\noindent
Witt, H.~J., \& Mao, S.\ 1994, \apj, 430, 505 

\begin{figure*}
\begin{center}
\psfig{file=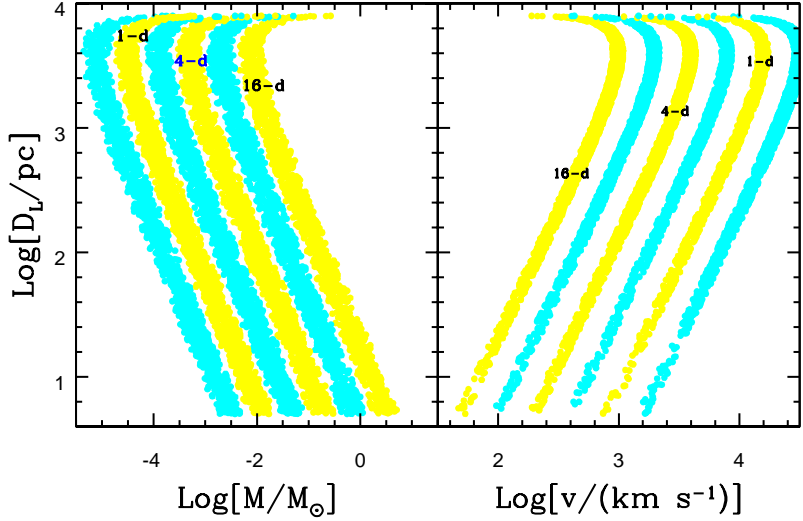,
height=6.0in,width=4.5in,angle=-0.0}
\vspace{-2.1 true in} 
\caption{
{\bf Left panel:} 
Distance to the lens, $D_L$, versus lens mass $M$. Each distinct swath
corresponds to events within a narrow range of Einstein diameter
crossing times, $\tau_{E}$.  The values of $\tau_{E}$
are within $10\, \%$ of a central value.
The left-most cyan swath corresponds to $\tau_E = 0.5$~days,
with the central value of $\tau_E$ doubling in each successive swath. 
The transverse speed was
taken to be $75$ km s$^{-1}$. 
{\bf Right panel:} 
Distance to the lens, $D_L$, versus the relative transverse
velocity, $v$. The lens mass $M$ is taken to be $1.4\, M_\odot,$
comparable to the mass of a neutron star.
Neutron stars with high kick velocities may be the most numerous
high-velocity objects.
Each distinct swath
corresponds to events within a narrow range ($10\%$) of Einstein diameter
crossing times, $\tau_{E}$:  $16$~days for the leftmost swath, decreasing by
a factor of $2$ for each swath to the right.
}
\end{center}
\end{figure*}

\begin{figure*}
\begin{center}
\psfig{file=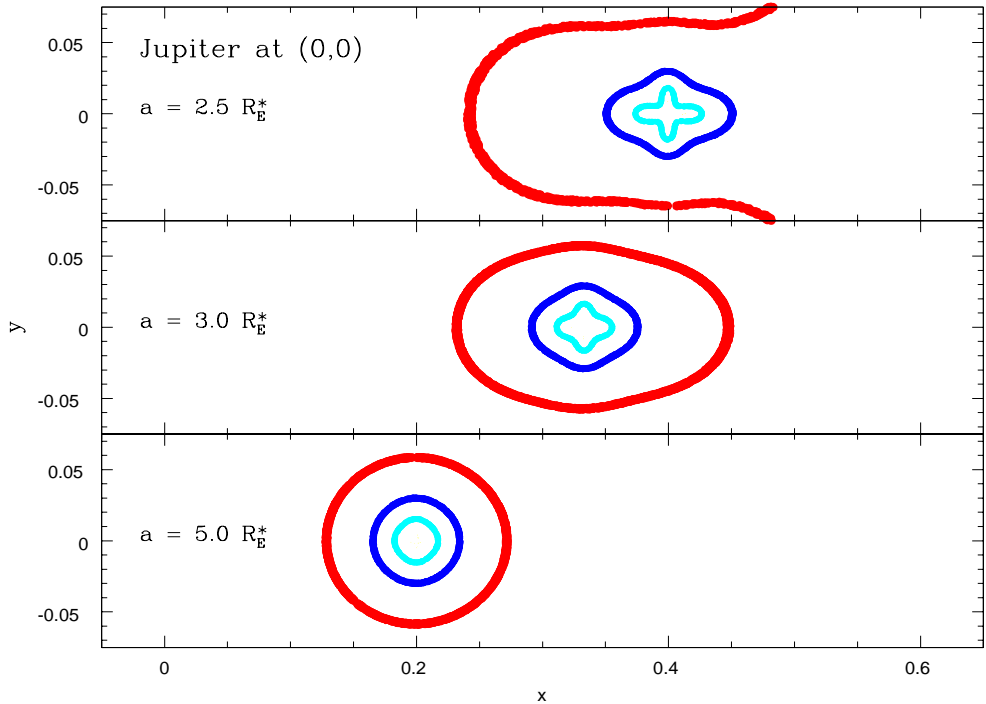,
height=5.5in,width=4.0in,angle=-90.0}
\caption{Magnification patterns associated with a planet lens with $q=0.001.$
Red: the magnification lies within $1\%$ of the value
expected for a point lens when $u=2\, \theta_E.$
Blue: the magnification lies within $1\%$ of the value
expected for a point lens when $u = 1\, \theta_E.$
Cyan:  the magnification lies within $1\%$ of the value
expected for a point lens when $u = 0.5\, \theta_E.$  
{\bf Bottom:} 
the planet is far enough from the star ($a=5\, R_{E,\ast}$)
that the isomagnification contours are symmetric
and show no evidence of the presence of the central star.
{\bf Middle:} the planet is closer to the star and
 the isomagnification 
contours are distorted; many short events generated by
the planet would 
require fit parameters modeling the relative mass and separation of the  
central star. {\bf Top:} The planet-star separation is $1.5\, R_{E,\ast}$
and 
the isomagnification contours are highly distorted; almost
every event generated by the planet would provide evidence of the central 
star.     
}
\end{center}
\end{figure*}

\end{document}